\documentstyle[11pt,newpasp,twoside,epsf]{article}
\begin{document}
\title{LIFTS: an Imaging Fourier Transform Spectrograph for Astronomy}
\author{Ron Wurtz\altaffilmark{1}, Edward Wishnow\altaffilmark{1}, S\'ebastien Blais-Ouellette\altaffilmark{2}, Kem Cook\altaffilmark{1}, Dennis Carr\altaffilmark{1}, Fr\'ed\'eric Grandmont\altaffilmark{3}, Isabella Lewis\altaffilmark{1} \& Christopher Stubbs\altaffilmark{4}}

\altaffiltext{1}{Lawrence Livermore National Laboratory, Box 808, Livermore CA 94551}
\altaffiltext{2}{D\'epartement de physique, Universit\'e de Montr\'eal, Montr\'eal,
Qu\'ebec, Canada, H3C 3J7}
\altaffiltext{3}{D\'epartement de Physique, G\'enie Physique et Optique,
Universit\'e Laval, Qu\'ebec, Qu\'ebec, Canada, G1K 7P4 and ABB Bomem, Inc., Qu\'ebec, Qu\'ebec, Canada, G1K 9H4}
\altaffiltext{4}{Astronomy Department, University of Washington, Box 351580,
Seattle WA, 98195}

\begin{abstract}
We present the first astronomical observations of the Livermore Imaging Fourier
Transform Spectrograph for visible-band astronomy using the 3.5-meter Apache
Point Observatory.
\end{abstract}

\section{Introduction}

We present the first astronomical results of LIFTS (see Wurtz et
al. 2001 for details). The double output Michelson interferometer
allow the observation of a spectrum for each of the 1k x 1k pixels of
the combined images.

In a collimated beam, the incoming light hits a beam splitter and
follows two possible optical paths before recombining into two output
CCD cameras. All the photons are therefore contained in the sum of the
two images while the relative phase shift information is stored in
their difference. When stepping through a range of path length
differences, by moving one arm of the interferometer, one ends up with
the Fourier transform of the spectrum for each element of the images.

\subsection{The Unique Advantages of an IFTS}

By first principles, an FTS has a highly flexible resolution from
${\cal R} = 1$ to several thousands, only limited by the maximum
travel of the moving arm ( 1 cm in our case). Every pixel contains a
spectrum, and all the images can be co-added to obtain a very deep
image, called the ``panchromatic'' image.  Follow-up spectroscopy of
faint objects discovered in the panchromatic image can be directly
obtained from the datacube. The observer can even decide to increase
the spectral resolution in real-time by increasing the number of steps
to a higher path length difference. We stress that this technique does
not require preselecting objects or subregions in a few-arcminute
field to be fed to a spectrograph.

Compared to any dispersive system, where all the spatial and spectral
elements have to fit on a single CCD, the IFTS has a substantial
multiplex advantage. Not only is the full CCD available for
spatial information, but the spectral information is just limited by
the number of channels one is willing to take on a given field. This
multiplex advantage is proportional to the density of relevant pixels
in the image (including some amount of sky and calibration objects).

The IFTS must be distinguished from filter systems including imaging
Fabry-Perot (FP) spectrographs or tunable filters. An IFTS accept all
the light at all wavelengths. It only creates one interference between
the two ``branches'' of a photon and records the phase difference. On
the other hand, an FP creates multiple ``branches'' by multiple
reflections inside a cavity creating constructive interferences only
for a narrow band of wavelengths, rejecting most of the light. Also,
the passband is periodic and one order has to be selected, limiting
the spectral range to a few angstroms. F-P systems are better suited
for narrow band observations of faint extended objects as they cut all
the out-of-band noise to which IFTS would be sensitive.

\section{Observations with the Livermore IFTS}

We have had a total of seven clear bright-time half-nights on the APO
3.5-meter with the completed instrument.  Seeing ranged from 0.7 to 1.5
arcsec, so we binned to $512\time 512$ pixels so that our pixel size
was 0.28 arcsec.
\begin{figure}
\plotfiddle{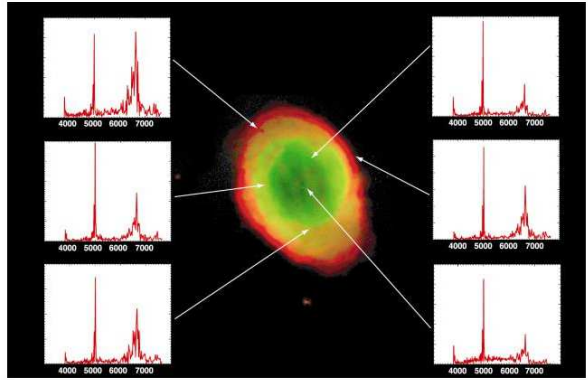}{5.7cm}{0}{40}{40}{-140}{-70}
\caption{ A view of a spectral-spatial datacube of the Ring Nebula,
obtained with LIFTS at Apache Point Observatory. The figure shows the
panchromatic image and the spectrum of representative pixels.}
\label{fig-2}
\end{figure}
The Ring Nebula datacube, at ${\cal R} \sim 430$, is presented
in Figure~\ref{fig-2}.

\end{document}